\documentclass[aps,preprint]{revtex4}%
\usepackage{amsfonts}
\usepackage{amsmath}
\usepackage{amssymb}
\usepackage{graphicx}%
\begin{document}
\preprint{ }
\title[Formulation of electrodynamics]{Fundamental formulation of electrodynamics revisited, and the precision of
quantum electrodynamics}
\author{H. R. Reiss}
\affiliation{Max Born Institute, Berlin, Germany}
\affiliation{American University, Washington, DC, USA}
\email{reiss@american.edu}

\begin{abstract}
It was shown recently that unambiguous description of electromagnetic
environments requires electromagnetic potentials; knowledge only of electric
and magnetic fields is insufficient and can lead to error. Consequences of
that demonstration are here applied to propagating fields, such as laser
fields. Gauge invariance is replaced by symmetry preservation. This alteration
makes it possible to understand how the known failure of the convergence of
perturbation expansions in quantum electrodynamics (QED) follows from the fact
that QED is incomplete; it does not contain its strong-field limit. Inherent
in that demonstration are the strong-field coupling constant and the
strong-field alteration of the mass shell of a charged particle. A variety of
physically important consequences ensue, including the loss of guidance from
Feynman diagrams. The meaning of tests for the precision of QED is questioned
since such evaluations apply only to perturbative QED, but not to extensions
required for complete QED.

\end{abstract}
\date[16 March 2020]{}
\maketitle

\section{Introduction}

Gauge transformations are designed to find alternative sets of electromagnetic
potentials associated with a specific configuration of electric and magnetic
fields. The conventional belief has long been that electric and magnetic
fields define the electromagnetic environment, so that the existence of
alternative sets of potential functions identifies them as auxiliary
quantities. That situation is inverted in Ref. \cite{hr100}, where gauge
transformations are shown to alter the fundamental symmetries associated with
a physical system. That opens the possibility of finding spurious solutions in
Maxwellian electrodynamics and Newtonian mechanics. It is necessary to select
an appropriate set of potentials to define fully the physical problem. This
identifies potentials as more fundamental than electric and magnetic fields.
It is also shown in Ref. \cite{hr100} that gauge transformations are not, in
general, unitary transformations.

The symmetries appropriate to a physical environment are determined by
properties of the Lagrangian describing the system \cite{noether}. The quantum
theory of fields is based on the properties of Lagrangians, and so that
discipline is free of the ambiguities explored here and in Ref. \cite{hr100}.
Gauge invariance is a foundation principle when employed in the quantum theory
of fields, but it has important limitations in the context of how electric and
magnetic fields are related to potential functions,

The conclusions of Ref. \cite{hr100} are applied to the important case where
the electromagnetic environment includes a propagating field (also known as a
transverse field, a plane-wave field, a sourceless field, or a photon field).
For example, these results apply to all laser-field problems.

The usual meaning of quantum electrodynamics\ (QED) is in the sense of
Feynman-Dyson perturbation theory, with the primary computational method being
the use of Feynman diagrams. The subjects examined here become very important
when strong fields exist, with the accompanying concepts of
intensity-dependent coupling constant and mass shell for charged particles.
These concepts are not part of standard QED. It is known that Feynman-Dyson
perturbation theory is not convergent. That lack is ascribed to the
incompleteness of QED, which does not include strong field phenomena.

The significance of studies of the precision of QED is questioned, since such
studies focus on the value of the fine-structure constant. That has meaning
only for perturbative QED, but not if the meaning of QED is extended to
include strong-fields. Laboratory experiments with strong fields interacting
with charged particles have for many years exceeded perturbative limits.
Importantly different conditions for extended QED are proposed.

The background for results presented here spans the entire history of
strong-field physics. It combines recent results, such as Ref. \cite{hr100},
with early work on strong fields from the 1950s and 1960s, whose significance
could not be fully appreciated at the time.

Gaussian units are used for electromagnetic quantities.

\section{Gauge invariance}

In the quantum theory of fields, gauge invariance has profound importance, but
that theory is entirely in terms of potentials. See, for example, Ref.
\cite{weinberg}. The sense in which gauge invariance is discussed here is in
the more mundane connection between potentials and electric and magnetic
fields. This semiclassical connection has not had a detailed examination
equivalent to that for quantized fields.

The principle of gauge invariance in its classical sense has, as its origin,
the notion that electric and magnetic fields determine all dynamical
consequences in electromagnetism, and that scalar and vector potentials have
only an auxiliary function. This type of gauge invariance is upset as a basic
principle when potentials are shown to be necessary to define an
electromagnetic environment. The propagation property constitutes a symmetry
that has not been adequately considered, but it is fundamental in showing that
there exist nominally valid\ gauge transformations that violate that symmetry.
Gauge invariance is replaced by symmetry preservation as a requirement for equivalence.

\subsection{Equations of quantum mechanics}

Acting upon the long-standing assumption that fields are basic and potentials
are secondary, many investigators attempted to express the Schr\"{o}dinger
equation directly in terms of fields
\cite{mandel,dewitt,belin,levy,rohrlich,priou}. All such attempts resulted in
making the Schr\"{o}dinger equation nonlocal and thus unsatisfactory. This
conclusion applies to relativistic equations of motion (Klein-Gordon, Dirac,
Proca) in addition to the nonrelativistic Schr\"{o}dinger and Pauli equations.
The implication, not realized at the time, is that potentials are more
fundamental than fields.

\subsection{Aharonov-Bohm effect}

The Aharonov-Bohm effect \cite{es,ab} is a direct demonstration that
potentials are more fundamental than fields in that the deflection of an
electron beam passing over a solenoid takes place in a region that is free of
fields, but has a potential that explains the deflection. This is a quantum
effect, and it is discussed in textbooks as being exclusively a quantum
anomaly that represents a departure from the notion that fields are primary
and potentials are secondary. Even that limited role has been questioned
\cite{jacksonokun}.

The Aharonov-Bohm effect involves a magnetic field, so it has no direct
significance for the study of propagating fields.

\subsection{Altered symmetries}

Reference \cite{hr100} demonstrates that a gauge transformation can alter the
basic symmetries that characterize a problem in electromagnetism. Symmetries
determine conservation properties \cite{noether}, so that a change in
symmetries represents a change to a different problem in electrodynamics. This
finding is quite general, and it occurs in both classical and quantum physics.
Different gauges for a given field configuration need not be equivalent, and
potential functions are required to define the appropriate electromagnetic environment.

A practical example presented in Ref. \cite{hr100} has important implications.
A propagating electromagnetic field, such as a laser field, must satisfy the
Einstein Principle \cite{einstein} that the speed of light in vacuum is the
same in all inertial frames of reference. The formal statement of this
principle is that the spacetime 4-vector $x^{\mu}$ can occur only as a scalar
product with the propagation 4-vector $k^{\mu}$. That scalar product:
\begin{equation}
\varphi\equiv k^{\mu}x_{\mu}=\omega t-\mathbf{k\cdot r}, \label{a00}%
\end{equation}
is the phase of a propagating field. This will be referred as the
\textit{propagation condition}. When the field is a propagating field, the
4-vector potential must be expressible as $A^{\mu}\left(  \varphi\right)  $.

A gauge transformation\ in electrodynamics is expressed as%
\begin{equation}
A^{\mu}\longrightarrow\widetilde{A}^{\mu}=A^{\mu}+\partial^{\mu}\Lambda,
\label{a0}%
\end{equation}
where $\Lambda$, the generating function for the transformation, is a scalar
function that satisfies the homogeneous wave equation:
\begin{equation}
\partial^{\mu}\partial_{\mu}\Lambda=0. \label{A}%
\end{equation}
When those conditions are satisfied, the electric and magnetic fields are
unchanged by the transformation.

A valid gauge transformation that produces an invalid 4-vector potential for a
transverse field has the generating function \cite{hr79,hr100}%
\begin{equation}
\Lambda=-A^{\mu}x_{\mu}. \label{b0}%
\end{equation}
This is a scalar function that satisfies the condition (\ref{A}) required of a
generating function and it is also stated covariantly. However, it is clear
that this will introduce a violation of the propagation condition because
$x^{\mu}$ appears in isolation from $k^{\mu}$. The gauge-transformed 4-vector
potential is
\begin{equation}
\widetilde{A}^{\mu}=-k^{\mu}\left(  x^{\nu}A_{\nu}^{\prime}\right)  ,\text{
where }A_{\nu}^{\prime}\equiv\frac{d}{d\varphi}A_{\nu}\left(  \varphi\right)
. \label{a}%
\end{equation}
The propagation condition is violated, but the electric and magnetic fields
are unchanged by this gauge transformation. This violation of a basic property
of a propagating field leads to the general conclusion that potentials are
more fundamental than fields, since preservation of the fields can
nevertheless lead to an invalid representation of a propagating wave.

An alternative form of the gauge-transformed potential in Eq. (\ref{a}) is
\cite{hr79,hr100}%
\begin{equation}
\widetilde{A}^{\mu}=-\left(  \frac{k^{\mu}}{\omega/c}\right)  \mathbf{r\cdot
E}\left(  \varphi\right)  . \label{b}%
\end{equation}
Although this gauge is unacceptable, it has found some favor \cite{heidelberg}
because of its resemblance to the ubiquitous length-gauge potential
$-\mathbf{r\cdot E}\left(  t\right)  $.

\subsection{Gauge invariance versus symmetry preservation}

The historically accepted conditions required to maintain electric and
magnetic fields in a gauge transformation are insufficient to characterize an
electromagnetic environment. For maintenance of the propagation property of a
plane-wave field, a necessary additional requirement is preservation of the
propagation condition of Eq. (\ref{a00}).

When an electromagnetic environment has both a propagating field and a scalar
field,\thinspace\ then only the radiation gauge (also known as Coulomb gauge)
can be compatible with an origin of coordinates, typifying a scalar potential
such as a Coulomb field and, simultaneously, the absence of an origin of
coordinates necessary to describe a propagating field such as a laser field
\cite{hrjpb50}. In the radiation gauge, the time component of the 4-vector
potential represents the scalar field, and the 3-vector component represents
the propagating field. When only the propagating field is present, the Lorenz
condition $\partial^{\mu}A_{\mu}=0$ reduces to $\mathbf{\nabla\cdot A}=0$ .
The 3-vector gradient condition is often stated to be the defining condition
for the radiation gauge.

The general expression for a gauge transformation is given by Eq. (\ref{a0}).
The condition that $A^{\mu}$ must be a function only of $\varphi$ imposes the
same constraint on $\Lambda$, leading to%
\begin{equation}
\widetilde{A}^{\mu}=A^{\mu}+k^{\mu}\Lambda^{\prime}, \label{d}%
\end{equation}
where $\Lambda^{\prime}=d\Lambda/d\varphi$. That is, the only alteration of
the potential that is possible differs from $A^{\mu}$ by a component that lies
on the light cone. An important consequence is that%
\begin{equation}
\widetilde{A}^{\mu}\widetilde{A}_{\mu}=A^{\mu}A_{\mu}. \label{e}%
\end{equation}
This follows from the transversality condition $k^{\mu}A_{\mu}=0$ as well as
the fact that a 4-vector on the light cone is self-orthogonal: $k^{\mu}k_{\mu
}=0.$

Gauge invariance employed in classical and semiclassical electrodynamics as a
general principle cannot be correct since a gauge change will normally alter
symmetry conditions, meaning that the physical problem is changed. For
radiation fields in interaction with matter, gauge invariance is replaced by
symmetry preservation.

There is some flexibility possible even when the propagation condition is
enforced. That possibility arises when the generator of the gauge
transformation itself satisfies the propagation condition, as expressed in Eq.
(\ref{d}). A further differentiation gives%
\begin{equation}
\partial_{\mu}\widetilde{A}^{\mu}=\partial_{\mu}A^{\mu}+k_{\mu}k^{\mu}%
\Lambda^{^{\prime\prime}}=\partial_{\mu}A^{\mu}, \label{e1}%
\end{equation}
where $\Lambda^{^{\prime\prime}}=d^{2}\Lambda/d^{2}\varphi$ and $k_{\mu}%
k^{\mu}=0$. The Lorenz condition $\partial_{\mu}A^{\mu}=0$ and the condition
of Eq. (\ref{A}) are automatically satisfied.

\subsection{Length gauge aberration}

Of the several gauges in use in strong-field physics, there is one that stands
out for the insupportable claims made for it.

The \textquotedblleft length gauge\textquotedblright\ is a name used to refer
to the $-\mathbf{r\cdot E}\left(  t\right)  $ scalar function to represent the
interaction of radiation fields with matter. There is an influential body of
literature devoted to the claim that the length gauge is the only proper gauge
to be used, and if a different gauge is to be employed, then it must always
carry with it the gauge transformation factor. This hypothesis is here termed
an \textquotedblleft aberration\textquotedblright\ since it is not possible
for a scalar potential to be fundamental for treatment of a vector field like
a laser field.

Two of the most ambitious efforts advocating the primacy of the length gauge
are cited here: \cite{yang,lss}. Both of these papers assume that the
interaction Hamiltonian behaves unitarily in a gauge transformation. This is
untrue \cite{hr100}. It is plainly a contradiction since $U\left(
\mathbf{r\cdot E}\right)  U^{-1}$ remains just $\mathbf{r\cdot E}$ in any
attempted gauge transformation because that interaction Hamiltonian contains
no operators. This means that, if a gauge transformation is made into the
length gauge, then it is impossible to do the inverse transformation back to
the initial gauge. This is untenable.

\section{Quantum electrodynamics is incomplete}

In 1952, Dyson showed \cite{dyson} that the perturbation expansion of QED is
not convergent. This continues to pose a dilemma since increasing accuracy in
both experiments and computation have failed to show any discrepancies. Dyson
conjectured that QED was somehow incomplete.

That Dyson's conjecture is correct was demonstrated long ago
\cite{hrdiss,hr62}. Standard QED has the basic defect that it does not contain
its strong-field limit. That is, QED exists as a perturbation expansion
without knowledge of the complete theory to which it is an approximation.

The identification of the defect followed from an effort to find the
convergence properties of relativistic quantum mechanics (RQM) as obtained
from a problem using the Volkov solution \cite{gordon,volkov}. This is an
exact solution of the Klein-Gordon equation found by Gordon \cite{gordon} for
a scalar charged particle in a plane-wave field, and an exact solution of the
Dirac equation found by Volkov \cite{volkov} for a spin-%
%TCIMACRO{\U{bd} }%
%BeginExpansion
${\frac12}$
%EndExpansion
particle in a plane-wave field. It has become conventional to refer to both
solutions as the Volkov solution. In the context of the application to
Breit-Wheeler pair production \cite{bw} for arbitrarily high intensities, an
analysis showed that RQM possesses a convergent perturbation expansion with
the radius of convergence limited by intensity-dependent singularities in the
complex coupling-constant plane. Two very important and unexpected features
arose in the investigation: an altered mass shell, and an intensity-dependent
coupling constant.

\subsection{Strong-field mass shell}

The quantity referred to as the \textquotedblleft mass shell\textquotedblright%
\ is the expression%
\begin{equation}
p^{\mu}p_{\mu}=\left(  mc\right)  ^{2}, \label{f}%
\end{equation}
where $p^{\mu}$ is the 4-momentum vector of the particle of mass $m$. It was
discovered independently by Sengupta \cite{sengupta} and by the present author
\cite{hrdiss,hr62} that, in strong fields, the mass shell is altered to%
\begin{equation}
p^{\mu}p_{\mu}=\left(  mc\right)  ^{2}+2mU_{p}, \label{g}%
\end{equation}
where $U_{p}$ is the ponderomotive potential of a particle of charge $q$ in a
transverse field, defined as%
\begin{equation}
U_{p}=\frac{q^{2}}{2mc^{2}}\left\langle \left\vert A^{\mu}A_{\mu}\right\vert
\right\rangle . \label{h}%
\end{equation}
The ponderomotive potential is plainly Lorentz invariant and, from Eq.
(\ref{e}) it is also gauge-invariant for a propagating field. The absolute
value $\left\vert A^{\mu}A_{\mu}\right\vert $ is employed because $A^{\mu}$ is
a spacelike 4-vector and it is best to use a positive number as a basic
measure. The angle brackets in Eq. (\ref{h}) refer to an average over a full
cycle of the field. That is employed because, within any cycle of oscillation
in a periodic field, it is known \cite{ss} that there is a continuous exchange
between kinetic and potential energies even though, in any complete cycle,
there can be no net energy transfer between a transverse field and a free
charged particle.

The difference between the mass shell expressions in Eqs. (\ref{g}) and
(\ref{h}) is minor in experiments with low-power laser beams, but it is a
major factor with modern high-power pulsed laser beams.

\subsection{Strong-field coupling constant}

The coupling constant in standard QED between a plane-wave field and a
particle of charge $e$ is the fine-structure constant $\alpha=e^{2}/\hslash
c$. In the convergence investigation of Refs. \cite{hrdiss,hr62} it was found
that the coupling parameter\ of strong-field physics is given by the
dimensionless intensity-dependent quantity%
\begin{equation}
z_{f}=2U_{p}/mc^{2}. \label{i}%
\end{equation}
Expressed in terms of $z_{f}$, the mass shell of Eq. (\ref{g}) is%
\begin{equation}
p^{\mu}p_{\mu}=\left(  mc\right)  ^{2}\left(  1+z_{f}\right)  , \label{j}%
\end{equation}
so that $z_{f}$ is a direct measure of the distinction between strong-field
and standard electrodynamics.

There is another way to express $z_{f}$ that is very informative. If $\alpha$
is extracted as a multiplier, then $z_{f}$ can be written in the form
\cite{hrrev,hr89}%
\begin{equation}
z_{f}=\alpha\rho V=\alpha\rho\left(  2\lambdabar_{C}^{2}\lambda\right)  ,
\label{k}%
\end{equation}
where $\rho$ is the density of photons and $V=2\lambdabar_{c}^{2}\lambda$ is
the volume that supplies photons to a strong-field process. This volume is
essentially a cylinder of radius $\lambdabar_{C}$ and length $\lambda$, where
$\lambdabar_{C}$ is the Compton wavelength and $\lambda$ is the wavelength of
the propagating field. The Compton wavelength is the usual measure of the
interaction radius of a charged particle in a propagating field. The
wavelength $\lambda$ is a macroscopic quantity. That is, all the photons in
this cylinder contribute to the interaction even though $\lambda$ might be
many orders of magnitude larger than the size of a target that is subjected to
the strong field.

It is the presence of a macroscopic quantity that can be said to characterize
a strong field.

\subsection{Dressed electrons}

When an electron (or any charged particle) is immersed in a strong propagating
field, it possesses a field-caused potential energy of $U_{p}$. This energy
comes from the propagating field, which is a relativistic phenomenon, so that
the electron must also acquire the momentum $U_{p}/c$ of the \textquotedblleft
dressing\textquotedblright\ field. Photons from the background field have a
4-momentum on the light cone. That is, the electron must acquire the
4-momentum%
\begin{equation}
U^{\mu}=U_{p}\left(  \frac{k^{\mu}}{\omega/c}\right)  . \label{k1}%
\end{equation}
The dressed electron can be regarded as a free particle with the 4-momentum
$p^{\mu}+U^{\mu}/c,$ and satisfy the usual mass shell condition of Eq.
(\ref{f}), which becomes%
\begin{equation}
\left(  p^{\mu}+\frac{1}{c}U^{\mu}\right)  \left(  p_{\mu}+\frac{1}{c}U_{\mu
}\right)  =p^{\mu}p_{\mu}+\frac{2}{c}p^{\mu}U_{\mu}, \label{k2}%
\end{equation}
since $U^{\mu}U_{\mu}\sim k^{\mu}k_{\mu}=0$. The Lorentz invariant quantity on
right-hand side of this equation can be evaluated in the rest frame of the
electron, so that Eq. (\ref{k2}) becomes%
\begin{equation}
\left(  p^{\mu}+\frac{1}{c}U^{\mu}\right)  \left(  p_{\mu}+\frac{1}{c}U_{\mu
}\right)  =\left(  mc\right)  ^{2}+2mU_{p}, \label{k3}%
\end{equation}
which is exactly Eq. (\ref{g}). That is, Eq. (\ref{g}) can be regarded as the
mass shell of a free electron dressed by the propagating field.

\subsection{Summed Feynman diagrams}

Fried and Eberly \cite{zfje} showed that it was possible to sum the Feynman
diagrams of QED to all orders in a Compton scattering problem in which the
spinor electron is replaced by a scalar particle. With one important revision,
the result they found exactly duplicates what is obtained by using a Volkov
solution. The only difference is that $z_{f}$ does not appear, and the mass
shell of Eq. (\ref{f}) is obtained.

The Fried and Eberly calculation verifies that QED is incomplete.

A subsequent investigation by Eberly and Reiss \cite{jehr} examined a class of
diagrams, each of which is divergent, that was omitted from the Fried and
Eberly calculation on the grounds that the divergent diagrams are unphysical.
The Eberly and Reiss paper showed that these diagrams can be summed exactly,
and the sum is finite. The importance of this demonstration is that the summed
divergent diagrams introduce exactly the strong-field mass-shell expression of
Eq. (\ref{g}) or (\ref{j}).

The Eberly and Reiss calculation shows the manner in which QED fails to be complete.

\subsection{Radius of convergence of perturbation theory and the failure of
Feynman diagrams}

The convergence investigation of Refs. \cite{hrdiss} and \cite{hr62} shows
that an extreme upper limit for perturbation theory is marked by the first
channel closing that can occur. A feature of strong-field processes is that
the field must supply the basic energy required for a transition as well as
the potential energy $U_{p}$ of any created charged particle immersed in a
strong field. For example, in the strong-field Breit-Wheeler pair production
process, multiple photons are needed to supply the rest energy $2mc^{2}$ of
the pair produced, but also the ponderomotive energy of the two charged
particles created. As the field intensity increases, the lowest-order process
that can occur must index upward to supply the required ponderomotive energy
of the electron pair. That indexing is referred to as a \textquotedblleft
channel closing\textquotedblright, and perturbation theory will fail at or
before the intensity for the first channel closing.

Reference \cite{jehr} shows that an infinite number of Feynman diagrams must
be summed in order to explain the strong-field mass shell. The strong-field
mass shell is therefore a nonperturbative effect. When the intensity is high
enough for perturbation theory to fail, then Feynman diagrams lose all meaning
since they provide a pictorial representation only of perturbative processes.

Figures 7 and 8 of Ref. \cite{hr80} show graphically the change from
perturbative to nonperturbative behavior. At low field intensity the lowest
allowed order of interaction is the sole contributor to a transition. At high
field intensity the superposition of many different photon orders is necessary
to describe a quantum transition.

\subsection{Green's function of the Volkov solution}

The properties of the Green's function of the Volkov solution provide a clear
picture of the differences between a standard propagator in QED and that for
strong fields. It is conventional to examine the behavior of the Green's
function in a complex $p^{0}$ space (i.e. complex energy space). This is
instructive because the path employed for an integration in this space
establishes whether the Green's function represents an advanced solution, a
retarded solution, or a Feynman solution in which positive energy solutions
propagate forward in time and negative energy solutions propagate backward.

In QED, the mass shell of Eq.(\ref{f}) applies, and the only poles in the
complex energy space occur at $p^{0}=\pm\sqrt{\mathbf{p}^{2}+mc^{2}}$. In the
Volkov Green's function, the mass shell of Eq. (\ref{g}) applies, and families
of sideband singularities appear in addition to the two QED poles. Each QED
pole has its own set of sideband states, but they do not overlap. See Ref.
\cite{hrje} for details.

These Green's function properties make explicit the differences found between
the QED calculation of Ref. \cite{zfje} and the strong field calculation of
Ref. \cite{jehr}. QED has but two singularities on the complex $p^{0}$ space,
whereas the strong-field case has infinite families of sidebands. This
phenomenon illustrates the failure of Feynman diagrams, which cannot represent
an infinity of singularities.

\section{Strong-field inferences}

Two important strong-field matters will be mentioned here, in addition to the
strong-field features discussed in preceding Sections.

\subsection{Fixed origin for energy measure}

The ponderomotive energy $U_{p}$ is the fundamental measure applicable to a
charged particle in a strong field. This is reflected in the essential
properties of $z_{f}$ as the coupling constant for charged particle
interactions with strong fields, as well as its role in the strong-field
altered mass shell. When interactions of the field with bound-state particles
are considered, another dimensionless intensity parameter arises, which is the
ratio of the ponderomotive potential to the binding energy $E_{B}$
\cite{hr80,hrrev}%
\begin{equation}
z_{1}=2U_{p}/E_{B}. \label{o}%
\end{equation}
A third dimensionless parameter is
\begin{equation}
z=U_{p}/\hslash\omega, \label{p}%
\end{equation}
which is of universal applicability in strong-field problems since it is a
measure of the minimum number of photons that enter into a
strong-field-induced interaction. As $U_{p}$ increases, this minimum indexes
upward, illustrating channel closing. As mentioned above, the first such
channel closing identifies an upper limit for the convergence of perturbation
theory \cite{hrdiss,hr62,hr80,hrrev}.

A novel feature that follows from the basic importance of $U_{p}$ is that it
fixes an absolute origin for energy measures. As Eq. (\ref{h}) shows, the zero
of energy is established by the zero of the 4-vector $A^{\mu}$. In the dipole
approximation as employed in atomic physics, the zero of energy can be
arbitrary as long as it is applied universally.

This fixed origin of energy measure is an important feature distinguishing the
two varieties of the Strong-Field Approximation (SFA).

\subsection{Ambiguity in the SFA}

The SFA is regarded as the standard analytical approximation for the
interaction of strong laser fields with matter. There is an existential
problem with this appraisal in that the SFA exists in two incompatible forms.

\subsubsection{SEFA}

The dipole approximation as used in atomic physics neglects entirely the
magnetic component of laser fields. When the dipole approximation is imposed
from the outset, the SFA is a theory of oscillatory electric fields. It is not
a theory of propagating fields like those of lasers. Despite the similarities
in some ranges of parameters, the differences are fundamental, and of major
importance in other ranges of parameters.

The first strong-field analytical approximation employed for laser-induced
processes is that of Keldysh \cite{keldysh}, who employs the dipole
approximation. This and subsequent dipole-approximation methods will be termed
the Strong Electric-Field approximation (SEFA).\newline

The dipole approximation is also employed in numerical solution of the
time-dependent Schr\"{o}dinger equation (TDSE), so that it is also of the SEFA
character. It is not an exact calculation of laser-induced transitions, as is
often claimed.

\subsubsection{SPFA}

A different analytical approximation follows from taking the nonrelativistic
limit of a theory based on propagating fields, which will be referred to as
the Strong Propagating-Field Approximation (SPFA). The genesis of the SFA of
Ref. \cite{hr80} from relativistic origins is demonstrated in Refs.
\cite{hr90,hrrel}. When a laser field is very strong, it is the dominant
influence in interactions of the field with matter. Laser fields propagate at
the speed of light, so that a relativistic treatment is necessary. When such a
theory is reduced to its nonrelativistic long-wavelength form \cite{hr80}, its
provenance from a relativistic formalism remains important even though the
general appearance of the SPFA resembles that of the SEFA.

\subsubsection{SEFA/SPFA differences}

When field frequencies are relatively high, SEFA and SPFA theories coalesce.
This is shown in detail in Ref. \cite{bondar}, for example. The authors do
numerical integration of the time-dependent Schr\"{o}dinger equation,
employing the dipole approximation, so it corresponds to a SEFA. The method
they cite as SFA is the SPFA.

When field frequencies are low, then SEFA and SPFA predictions become
profoundly different \cite{hr101,hrtun}. The SEFA trends toward what has been
labeled as the \textquotedblleft asymptotic limit\textquotedblright, where the
field becomes a constant electric field. By contrast, as the field frequency
decreases, the SPFA increasingly manifests the effects of the magnetic
component of a laser field, trending towards relativistic behavior. The
location of the transition from high and low frequency domains has yet to be
established, but it corresponds approximately to wavelengths in the few-$\mu
m$ range.

A very important matter is that the SPFA of Refs. \cite{hr80,hr90,hrrel} is
the only strong-field approximation method that can be categorized as SPFA.
Everything else, including TDSE, is SEFA.

\section{Varieties of electrodynamics}

Electrodynamics can be viewed from the standpoint of quantum field theory
(QFT), or of relativistic quantum mechanics, or as a purely classical
phenomenon. Each of these viewpoints overlaps the adjacent one, generating a
unified view of electrodynamics.

\subsection{Electrodynamics as a quantum field theory}

QFT has become a highly developed formalism based on symmetry principles,
leading to the modern \textquotedblleft Standard Model". Electrodynamics has a
place in this scheme as the gauge particle of the electromagnetic field. For
purposes of this article, the discipline that earned a Nobel prize for
Feynman, Schwinger, and Tomonaga is sufficient. The salient point in QFT is
the existence of a number operator whose eigenvalues count the number of
photons that participate in an interaction.

The practical application of quantum electrodynamics to laboratory phenomena
is accomplished according to the graphic means of the Feynman diagrams that
follow from perturbation theory.

\subsection{Electrodynamics in relativistic quantum mechanics}

In RQM, the field is not quantized. That is, there is no number operator.
Nevertheless, RQM employs what is called the Floquet property, in which the
periodicity of an electromagnetic plane wave leads to transfer of energy in
integer packets of $\hslash\omega$. Using standard S-matrix methods in a
relativistic formulation, a set of computational rules can be evolved that are
identical to the Feynman rules of QFT. A clear representation of the equality
of the Feynman rules in QED and in RQM is given by the two textbooks of
Bjorken and Drell \cite{bd1,bd2}. The first volume uses RQM methods to produce
the Feynman rules, followed by the QFT volume that produces exactly the same rules.

The novel feature of RQM is the existence of the Volkov solution, which makes
possible the construction of a nonperturbative domain of electrodynamics. The
essential distinctions between the QFT of electrodynamics and electrodynamics
within RQM is elucidated by Refs. \cite{zfje} and \cite{jehr}.

\subsection{Classical electrodynamics}

Classical electrodynamics does not employ a quantized version of the
electromagnetic field, but it is nevertheless possible to define the photon
density of a monochromatic field by using the classical energy density of a
plane-wave field divided by $\hslash\omega$. This is the method used in Eq.
(\ref{k}) to evaluate photon density.

The application of classical electrodynamics to such practical matters as
antenna theory or the properties of transmission lines seems to have no
correspondences with RQM or QFT, but there is nevertheless an important
connection with nonperturbative RQM. This connection arises through the
wavelength dependence of the coupling constant of RQM in the form given by Eq.
(\ref{k}). The salient question arises from the possibility that $\lambda$ can
be such a large quantity that its connection to the microscopic world of
quantum mechanics becomes difficult to understand.

Some insight into this question comes from a situation in which classical
phenomena at extreme wavelengths is also difficult to understand.

There is a practical application of extremely long wavelengths to the problem
of communication with deeply submerged submarines. Several countries have
devised systems operating at very long wavelengths to take advantage of the
fact that the skin depth of a conducting medium such as seawater varies a
$\lambda^{1/2}$. The system employed by the U.S. Navy \cite{wiki} operated at
a frequency of $76Hz$, corresponding to a wavelength of $4\times10^{6}m$,
which is almost equal to the Earth's radius of $6.4\times10^{6}m$. The
receiving antenna can be regarded as the length of the submarine, of the order
of $10^{2}m$, or one part in $40,000$ of the wavelength of the radio signal.
On the scale of the submarine, the electric and magnetic fields of the radio
wave are constant. Nevertheless, the radio wave carries a coded signal that is intelligible.

This is related to the problem of constant crossed fields. The two
relativistic invariants of a propagating electromagnetic field have zero
value:%
\begin{equation}
\mathbf{E}^{2}-\mathbf{B}^{2}=0,\quad\mathbf{E\cdot B}=0. \label{q}%
\end{equation}
It is also possible to generate constant $\mathbf{E}$ and $\mathbf{B}$ fields
that satisfy the conditions (\ref{q}). If the electric and magnetic fields are
the governing quantities that identify fields, then constant crossed fields
and propagating fields of very long wavelengths should be equivalent. They are
not. Transverse fields propagate in vacuum at the speed of light. Constant
crossed fields \textquotedblleft propagate\textquotedblright\ at zero speed.
When identified by potentials, constant crossed fields and propagating fields
are unrelated. This is direct proof of the primacy of potentials over fields.

The problem of how an atom can respond to a propagating field many orders of
magnitude greater than the size of the atom is analogous to the problem of how
a submarine can decipher a radio signal with a wavelength $40,000$ times the
length of the submarine. In each case the target of the plane wave can respond
to the information carried by the potential functions of the propagating
field. Properties of the $\mathbf{E}$ and $\mathbf{B}$ fields are secondary.

The critical strong-field parameter $z_{f}$ varies as the square of the
wavelength. It is possible to achieve $z_{f}=O(1)$ with commercial
radio-frequency equipment. That is, familiar classical environments can
exhibit certain strong-field effects of powerful lasers.

\subsection{Summary of the varieties of electrodynamics}

Descriptions of the effects of transverse fields are equivalent within QFT and
RQM within the domain of the validity of perturbation theory of QFT. RQM makes
available an analytical continuation of the effects of the transverse field
into a domain where QFT fails. Classical electromagnetism as applied to
macroscopic problems has no relevance to the microscopic world of quantum
systems, but nonperturbative RQM shares some of the important behavior of
macroscopic transverse fields at very long wavelengths.

\section{Precision of QED}

Appraisals of the precision of QED are based on the accuracy to which the
value of $\alpha$ can be determined \cite{gabrielse}. That is, the premise is
accepted that $\alpha$ measures the coupling of transverse fields to charged
particles. One intent of the present article is to show that Feynman-Dyson
perturbation theory applies only to a subset of electrodynamics. Strong fields
are neglected, and the coupling parameter of strong fields is $z_{f}$, not
$\alpha$.

The proposition is now made that the failure of QED to be convergent is
governed by the inability of QED to explain strong field phenomena. QED is not
a subset of strong-field physics, but rather it is an approximation to
strong-field physics. The importance of strong fields is measured by
$z_{f}=2U_{p}/mc^{2}$, and perturbation theory in the context of strong fields
has an absolute limit $z<1$, in terms of the intensity parameter $z$ of Eq.
(\ref{p}). That is QED is subject to the limit
\begin{equation}
z_{f}=\frac{2U_{p}}{mc^{2}},\quad z=\frac{U_{p}}{\hslash\omega}<1,\quad
\text{so that }z_{f}<\frac{2\hslash\omega}{mc^{2}}. \label{r}%
\end{equation}
For a typical laser photon energy of $1.5eV$, the limit given in Eq. (\ref{r})
is $z_{f}<6\times10^{-6}$. This is an extreme upper limit, and nonperturbative
behavior is known to exist at much smaller $z_{f}$ values. The best-known
manifestation of nonperturbative behavior is the above-threshold ionization
(ATI) effect, first observed by Agostini, \textit{et al. }\cite{agostini}. The
first successful match of a nonperturbative theory to experiment, reported in
Ref. \cite{hrjpb87}, was for a case where the peak $z_{f}$ was $z_{f}%
=8\times10^{-6}$, and this was clearly well into the nonperturbative domain.

\end{document}